\documentclass[12pt]{article}
\pdfoutput=1
\usepackage[DIV13]{typearea}
\usepackage{amsmath}
\usepackage{amssymb}
\usepackage{graphicx}
\usepackage{cite}
\usepackage{bbold}
\newcommand{\be}{\beta}

\def\be{\begin{equation}}
\def\ee{\end{equation}}
\def\beq{\begin{equation}}
\def\eeq{\end{equation}}
\def\bc{\begin{center}}
\def\ec{\end{center}}
\def\bea{\begin{eqnarray}}
\def\eea{\end{eqnarray}}

\newcommand{\La}{\Lambda}
\newcommand{\bs}{\boldsymbol}
\begin{document}
\begin{titlepage}
\vspace*{-1cm}
\phantom{hep-ph/***} 
\flushright
\hfil{DFPD-2013/TH/05}
\hfil{TUM-HEP-883/13}\\

\vskip 1.5cm
\begin{center}
\mathversion{bold}
{\LARGE\bf A Realistic Pattern of Lepton Mixing and\\ Masses from $S_4$ and CP}\\[3mm]
\mathversion{normal}
\vskip .3cm
\end{center}
\vskip 0.5  cm
\begin{center}
{\large Ferruccio Feruglio}~$^{a),b)}$,
{\large Claudia Hagedorn}~$^{a),b),c)}$\\[2mm]
{\large and Robert Ziegler}~$^{d),e)}$
\\
\vskip .7cm
{\footnotesize
$^{a)}$~Dipartimento di Fisica e Astronomia `G.~Galilei', Universit\`a di Padova,\\ Via Marzolo~8, I-35131 Padua, Italy
\vskip .1cm
$^{b)}$~INFN, Sezione di Padova, Via Marzolo~8, I-35131 Padua, Italy
\vskip .1cm
$^{c)}$~SISSA, Via Bonomea 265, I-34136 Trieste, Italy
\vskip .1cm
$^{d)}$~Physik-Department, Technische Universit\"at M\"unchen, 
\\
James-Franck-Strasse, D-85748 Garching, Germany
\\
\vskip .1cm
$^{e)}$~
TUM Institute for Advanced Study, Technische Universit\"at M\"unchen, \\
Lichtenbergstrasse 2a, D-85748 Garching, Germany
\vskip .5cm
\begin{minipage}[l]{.9\textwidth}
\begin{center} 
\textit{E-mail:} 
\tt{feruglio@pd.infn.it}, \tt{hagedorn@pd.infn.it}, \tt{robert.ziegler@ph.tum.de}
\end{center}
\end{minipage}
}
\end{center}
\vskip 1cm
\begin{abstract}
We present a supersymmetric model with the flavour symmetry $S_4 \times Z_3$ and a $CP$ symmetry which are broken to a $Z_3$ subgroup of the flavour symmetry 
 in the charged lepton sector and to $Z_2 \times CP (\times Z_3)$ in the neutrino one at leading order. This model implements
an approach, capable of predicting lepton mixing angles and Dirac as well as Majorana phases in terms of one free parameter.
This parameter, directly related to the size of the reactor mixing angle $\theta_{13}$, can be naturally of the correct order in our model.
Atmospheric mixing is maximal, while $\sin^2 \theta_{12} \gtrsim 1/3$. All three phases are predicted: the Dirac phase is maximal, whereas
the two Majorana phases are trivial. The neutrino mass matrix contains only three real parameters at leading order and neutrino masses
effectively only depend on two of them. As a consequence, they have to be normally ordered and the absolute neutrino mass scale and the sum of the neutrino masses are predicted. 
The vacuum of the flavons can be correctly aligned.
We study subleading corrections to the leading order results and show that they are small.

\end{abstract}
\end{titlepage}
\setcounter{footnote}{0}

\section{Introduction}
\label{intro}

The discovery of neutrino oscillations and the first information on lepton mixing angles has led to an intense research in this field.
The data could shed light on the flavour mystery and help to establish some organizing principle relevant for 
fermion masses and mixing parameters. Before the determination of the reactor mixing angle $\theta_{13}$ several flavour symmetries, usually broken in a particular way, have been proposed, see \cite{reviews}, 
which give rise to mixing patterns with vanishing $\theta_{13}$. A prime example is tri-bimaximal (TB) mixing \cite{TB} which was well compatible with the experimental data 
and can be derived from flavour groups such as $A_4$ \cite{A4TB} and $S_4$ \cite{S4TB}. The discovery of a relatively large reactor mixing angle \cite{theta13_exp}, however, requires most likely
additional ingredients beyond the presence of a small discrete flavour symmetry. The simplest assumption which can be made is to invoke
 sufficiently large corrections to the leading order (LO) pattern such that current data \cite{global_latest} can be reproduced, see e.g. \cite{AFM}.
A disadvantage of this Ansatz is that the predictability of such models is in general reduced, since several corrections are present. 
Discrete groups larger than $A_4$ or $S_4$ have also been considered in order to accommodate a non-vanishing $\theta_{13}$ \cite{Delta96_384}.
Another possibility is to apply less stringent requirements on the residual symmetries in the charged lepton and neutrino sectors \cite{GnuZ2_acc}. 
In this case $\theta_{13}$ becomes essentially a free parameter. However, correlations among the other mixing angles and $\theta_{13}$ could still provide a testable prediction.

The possible role of $CP$ and its ability to explain a non-vanishing reactor mixing angle and to constrain, at the same time, CP phases
 has been explored to some extent in the past. In such a case
 $CP$ acts non-trivially on flavour space \cite{general_CPV} and is thus dubbed generalized $CP$ transformation. A well-known example is given by the 
$\mu\tau$ reflection symmetry \cite{mutaureflection_HS,mutaureflection_GL}, which exchanges a muon (tau) neutrino with a tau (muon) antineutrino
 in the charged lepton mass basis. If this symmetry is imposed, the atmospheric mixing angle is predicted to be maximal, while $\theta_{13}$ is in general non-vanishing
 for a maximal Dirac phase $\delta$. This symmetry is preserved by the neutrino mass matrices of two recently proposed models \cite{simplest_nu_mass_S4,S4andCP_MN} which combine the 
 flavour symmetries $S_4$ and $A_4$, respectively, with $CP$. In \cite{simplest_nu_mass_S4} a non-supersymmetric extension of the Standard Model (SM) is considered in which neutrino
 masses arise from the type I seesaw mechanism. The flavour symmetry, that is augmented by a direct product of $Z_2$ symmetries, is broken spontaneously through the vacuum expectation
 values (VEVs) of gauge singlets. The neutrino mass spectrum allows for normal as well as inverted hierarchy. However, crucial aspects, such as the vacuum alignment, are only sketched and restrictions on couplings being 
 real or imaginary seem to be imposed by hand. The model in \cite{S4andCP_MN} is also non-supersymmetric and neutrino masses are generated through the type II seesaw mechanism. Neutrino masses are normally ordered. 
 The VEVs of multiple Higgs doublets and triplets break the flavour symmetry spontaneously. However, their potential is not discussed explicitly. Furthermore, the $\mu\tau$ reflection symmetry
 of the neutrino mass matrix seems to be accidental and the value of the solar mixing angle is not protected by a symmetry.
 \footnote{Several other approaches can be found in the literature in which the interplay of a discrete symmetry and $CP$ is studied.
 In \cite{geometrical_CPV} models with the flavour symmetry $\Delta (27)$ or $\Delta (54)$ are discussed in which so-called geometrical CP violation is realized, i.e. the potential
 allows for a configuration in which the (relative) phases of the VEVs of different scalar fields are fixed and take values independent of the parameters of the potential. 
 Subsequently, several aspects, such as the influence of non-renormalizable operators on the stability of this vacuum and ways to 
accommodate fermion masses and mixing, have been studied. In  \cite{CP_acc_symm_potential} supersymmetric (SUSY) extensions of the SM with the double-valued dihedral group
$Q_6$ (also called $D'_3$) are analyzed in which $CP$ is spontaneously broken so that the so-called SUSY CP problem is solved and certain predictions for fermion masses and mixing are made.
The models in \cite{CPV_Tprime} focus on the attempt to predict CP phases by making use of complex Clebsch Gordan coefficients that are associated with certain irreducible representations
of the group $T'$ ($T'$ is the double cover of the well-known group $A_4$) and by assuming at the same time real couplings in the Lagrangian and real VEVs.}

Taking these examples as starting point, we have considered an approach in which a discrete flavour group $G_f$ and a $CP$ symmetry are broken in a particular way
 such that the Pontecorvo-Maki-Nakagawa-Sakata (PMNS) mixing matrix depends only on one continuous parameter. In the presence of a flavour symmetry $G_f$ 
$CP$ transformations cannot be defined arbitrarily, but are subject to certain consistency conditions \cite{FHZ12,CPGf_HD,review_CP_gauge}.
If these are fulfilled, we can consider a theory with $G_f$ and $CP$ and can study the scenario in which residual symmetries $G_e$ and $G_\nu$ 
are preserved in the charged lepton and neutrino sectors, respectively. $G_e$ is assumed to be a subgroup of $G_f$,  while, generalizing the idea
of the $\mu\tau$ reflection symmetry, $G_\nu$ is taken to be the direct product of $CP$ and a $Z_2$ symmetry contained in $G_f$.
In \cite{FHZ12} we have shown, in a model-independent way, that such a breaking leads to a mixing pattern with only one free parameter and thus mixing angles as well as
CP phases are strongly correlated. We have furthermore shown that for $G_f=S_4$ several new mixing patterns, well compatible with current experimental data \cite{global_latest}, 
can be derived.

In the present paper we discuss a concrete model for leptons in which the above idea of the breaking of a flavour and a $CP$ symmetry is implemented. 
The model is SUSY and adopts the gauge symmetry, lepton superfields and Higgs multiplets of the  
Minimal SUSY SM (MSSM).\footnote{The mass of the scalar resonance found at LHC \cite{AtlasCMS,Djouadi} is presumed to be accommodated by some mechanism.} 
$G_f=S_4\times Z_3$ plays the role of the flavour symmetry. The $Z_3$ symmetry is employed in order to naturally generate the charged lepton mass hierarchy. 
We introduce additional degrees of freedom, flavons and driving fields, which are singlets under the gauge group and are necessary for the correct spontaneous breaking of $G_f$
 and $CP$. Two distinct sets of flavons are present whose VEVs contribute either to the charged lepton or to the neutrino mass matrix at LO. This separation is achieved by the $Z_3$ symmetry
 as well as with the help of an additional $Z_{16}$ group that is, in general, responsible for forbidding unwanted operators. The VEVs 
of the former set of flavons leave $G_e$ invariant, while those of the latter $G_\nu=Z_2 \times CP (\times Z_3)$. The $Z_3$ symmetry preserved in the neutrino sector is irrelevant as regards
lepton mixing and masses and thus will be omitted in the following.
The vacuum alignment is achieved via the requirement of vanishing F-terms of the driving fields.  Neutrino masses arise in our model from the Weinberg operator.
The lepton mixing matrix is of the same form as in one of the cases exemplified in \cite{FHZ12} and the following predictions are obtained: $\sin^2 \theta_{23}=1/2$, $\sin^2 \theta_{12} \gtrsim 1/3$
and $\delta$ maximal. The free parameter which is directly related to the reactor mixing angle 
is naturally of the correct size due to its origin from a subleading contribution to the neutrino mass matrix. All mixing angles agree
within $3 \, \sigma$ with the latest global fit results \cite{global_latest}. Both Majorana phases are trivial at LO. The maximality of the atmospheric mixing angle and of the Dirac phase
are consequences of the invariance of the neutrino mass matrix under the $\mu\tau$ reflection symmetry in the charged lepton mass basis.
Since some of the contributions to the neutrino mass matrix that are compatible with the preservation of $G_\nu$ are suppressed in our model, the neutrino mass spectrum
can only be normally ordered. Indeed, the three neutrino masses effectively depend on only two different parameters so that, after fitting the two known mass squared differences,
the absolute mass scale and the sum of the neutrino masses are strongly constrained.

The paper is organized as follows: in section 2 we briefly recall the framework of \cite{FHZ12} and we apply it to the case of our interest. We assign the transformation properties of  
lepton and Higgs supermultiplets under $G_f=S_4\times Z_3$ and $CP$ and we determine the most general VEVs of the flavon fields, compatible with the desired residual symmetries in the charged lepton
and neutrino sectors. In section 3 we specify the superpotential involving lepton superfields, we analyze charged lepton and neutrino mass matrices, both at LO and at
next-to-leading order (NLO), and we present the results for lepton masses and mixing parameters. In section 4 we discuss the vacuum alignment
and show that at LO the desired pattern of VEVs is a minimum of the scalar potential in the SUSY limit.
Furthermore, we examine NLO contributions to the flavon superpotential which induce shifts in the flavon VEVs. 
In section 5 we  summarize the main results of our paper. Some details of the group theory of $S_4$ are found in appendix A.

\section{Approach}
\label{sec2}

We elaborate on a proposal which we have recently presented in \cite{FHZ12}. We assume a finite, discrete flavour group, here $G_f=S_4 \times Z_3$, the direct product of the permutation
group of four distinct objects and the cyclic symmetry $Z_3$. This group is combined with a generalized $CP$ transformation $X$ which also acts on flavour space. If these 
symmetries are broken in a particular
way, all lepton mixing angles and CP phases can be predicted in terms of only one free parameter.

The group $S_4$ can be conveniently represented in terms of three generators $S$, $T$  and $U$ with $S$ and $U$ having order two, while
$T$ is of order three \cite{S4generators}. For details about the group theory of $S_4$ see appendix \ref{appA}.
A generic multiplet of fields is denoted by $\phi$ and transforms in a certain representation $({\bf r}, q)$ of  the flavour group $G_f$ 
(${\bf r}$ is one of the $S_4$ representations and $q$ stands for the $Z_3$ charge of the multiplet: 1, $\omega$ or $\omega^2$ with $\omega=e^{2 \pi i/3}$).
The $CP$ transformation $X$ acts on $\phi$ as
\begin{equation}
\label{dCP}
\phi' (x) = X \phi^* (x_{CP})
\end{equation}
with $x_{CP}=(x^0,-\vec{x})$. The choice of $X$ is not
arbitrary, but $X$ has to fulfill certain conditions: first, we require $X$ to be a unitary and symmetric matrix,
\begin{equation}
\label{Xcond1}
X X^\dagger = X X^* = \mathbb{1} \; ,
\end{equation}
so that $CP^2=1$ holds. Secondly, $X$ also has to fulfill a consistency condition involving
transformations which belong to the group $G_f$. For all elements $g$ of the group $G_f$ the following statement has to hold:
let $A$ be the (unitary) matrix which represents $g$ in $({\bf r}, q)$, then an element $g'\in G_f$ should exist such that
\begin{equation}
\label{Xcond2}
(X^{-1}A X)^*=A'
\end{equation}
with $A'$ denoting the (unitary) matrix associated with $g'$ in $({\bf r}, q)$.
In general, $g$ and $g'$ are distinct elements.
The mathematical structure of the group $G_{CP}$ comprising $G_f$ and $CP$ turns out to be a semi-direct product and can be written as
$G_{CP}= G_f \rtimes H_{CP}$ with $H_{CP}$ being the parity group generated by $CP$ \cite{FHZ12, CPGf_HD,review_CP_gauge}.

Closely following the analysis in \cite{FHZ12}, we show that
the lepton mixing parameters can be predicted from the breaking of the group $G_{CP}$ to $G_\nu= Z_2 \times CP$ in the neutrino and to $G_e=Z_3^{(D)}$ in the charged lepton sector. 
The group $Z_3^{(D)}$ is the diagonal subgroup of the group $Z_3 \subset S_4$ that is generated by $T$ 
and the additional $Z_3$ symmetry.\footnote{Compare also \cite{HS11} for such a type of residual symmetry $G_e$.} 
The requirement that $G_\nu$ should be a direct product of $Z_2 \subset G_f$ and $CP$ leads to a further constraint on $X$
\begin{equation}
\label{Xcond3}
XZ^*-ZX=0
\end{equation}
with $Z$ being the generator of the $Z_2$ symmetry in the representation $({\bf r}, q)$. 

In \cite{FHZ12} we have performed a comprehensive study of all admissible $X$ for the possible choices of $Z$ in the case of $G_f=S_4$. As can be checked these transformations $X$ are also viable for $G_f=S_4 \times Z_3$, since
the cyclic symmetry $Z_3$ in $G_f$ does not impose any further constraints on $X$. In order to simplify our considerations regarding the diagonal subgroup $Z_3^{(D)}$, we choose a basis
in which the generator $T$ is diagonal in all representations of $S_4$. Its explicit realizations as well as those of the other two generators $S$ and $U$ can be found in appendix \ref{appA}. 

If we take
\begin{equation}
\label{choiceZ}
 Z=S \;\; ,
 \end{equation}
a consistent choice for $X$ satisfying eqs.(\ref{Xcond1}-\ref{Xcond3}) is 
\begin{equation}
\label{otherX}
X_{{\bf 1}} =1 \; , \; X_{{\bf 1'}} =-1 \; , \; X_{{\bf 2}}=\left( \begin{array}{cc}
    0&1 \\
    1&0
    \end{array} \right) \; , \;
X_{{\bf 3}}=
 - \left( \begin{array}{ccc}
    1 & 0 & 0 \\
    0 & 0 & 1 \\
    0 & 1 & 0
    \end{array}\right) \; , \;
    X_{{\bf 3'}} =  \left( \begin{array}{ccc}
 1 & 0 & 0\\
 0 & 0 & 1\\
 0 & 1 & 0
\end{array}
\right) \; .
\end{equation}
Here we denote by $X_{{\bf r}}$ the realization of the $CP$ transformation $X$ in the representation ${\bf r}$ of $S_4$. This choice together with
 eq. (\ref{dCP}) specifies the action of  $CP$ on flavour space. 
Note that the transformations in eq. (\ref{otherX}) correspond to the choice $X_{{\bf 3'}}=\mathbb{1}$ in \cite{FHZ12}
(this can be checked by applying the basis transformation mentioned in footnote 7 in \cite{FHZ12}). 
The generalized $CP$ transformation $X_{{\bf 3'}}$ in eq. (\ref{otherX}) is known as $\mu\tau$ reflection symmetry in the literature \cite{mutaureflection_HS,mutaureflection_GL}.  

Let us consider that left-handed leptons $l$ transform as $({\bf 3'},1)$ under $S_4\times Z_3$. For $G_e=Z_3^{(D)}$, the combination $m_l^\dagger m_l$ 
($m_l$ is the charged lepton mass matrix in the right-left basis, $l^c \, m_l \, l$) 
 is diagonal and thus does not contribute to the lepton mixing, up to possible relabeling of the generations. For neutrinos we assume that they are Majorana particles and 
 that their masses arise from the Weinberg operator. 
Preserving $G_\nu=Z_2 \times CP$ in the neutrino sector requires that the neutrino mass matrix $m_\nu$ fulfills the 
invariance conditions 
\be
\label{condmnu}
Z^Tm_\nu Z=m_\nu~~~,~~~~~~~X_{{\bf 3'}} m_\nu X_{{\bf 3'}}= m_\nu^*~~~
\ee
from which follows
\be
U_{PMNS}^T~m_\nu~ U_{PMNS}=m_\nu^{diag}~~~
\label{diagmnu}
\ee
with
\begin{equation}
\label{PMNSCP}
U_{PMNS}= \frac{1}{\sqrt{6}} \, \left( \begin{array}{ccc}
 2 \cos \theta & \sqrt{2} & 2 \sin \theta \\
 -\cos \theta + i \sqrt{3} \sin \theta & \sqrt{2} & -\sin \theta - i \sqrt{3} \cos \theta \\
 -\cos \theta - i \sqrt{3} \sin \theta & \sqrt{2} & -\sin \theta + i \sqrt{3} \cos \theta \\
\end{array}
\right) \; K
\end{equation}
and $m_\nu^{diag}$ containing the neutrino masses $m_i$, $i=1,2,3$.
The unitary matrix $K$ is diagonal with entries $\pm 1$ and $\pm i$ which encode the CP parity of the neutrino states.
Notice that, without a particular model, the mixing matrix is only determined up to permutations of rows and columns, since neither charged lepton nor neutrino masses are constrained by the above conditions. 
All mixing parameters, angles and Dirac as well as Majorana phases, are predicted in terms of
the parameter $\theta$, which depends on the entries of the neutrino mass matrix and that can take in general
values between $0$ and $\pi$. From eq.(\ref{PMNSCP}) the mixing angles can be read off\footnote{Our convention for mixing angles and CP phases can be found in appendix A.1 in \cite{FHZ12}.} 
\begin{equation}
\label{mixingangles}
\sin^2 \theta_{13} =\frac{2}{3} \sin^2 \theta \; , \;\; \sin^2 \theta_{12}=\frac{1}{2+\cos 2 \theta} \; , \;\; \sin^2 \theta_{23}=\frac{1}{2} \; .
\end{equation}
Thus, for $\theta \approx 0.185$ the best fit value of the reactor mixing angle can be accommodated well, $\sin^2 \theta_{13} \approx 0.023$, and also the solar and atmospheric mixing angles are within their $3 \, \sigma$ ranges \cite{global_latest},
$\sin^2 \theta_{12} \approx 0.341$. For a $\chi^2$
analysis we refer to \cite{FHZ12}. Since the second column of the mixing matrix in eq.(\ref{PMNSCP}) is tri-maximal, the solar mixing angle is constrained by $\sin^2 \theta_{12} \geq 1/3$ \cite{TM_pheno}. 
The Majorana phases $\alpha$ and $\beta$ are trivial,
\begin{equation}
\sin \alpha =0 \; , \;\; \sin \beta =0 \; ,
\end{equation}
while the Dirac phase $\delta$ is maximal
\begin{equation}
|\sin \delta|=1 
\end{equation}
and thus $|J_{CP}|=|\sin 2 \theta|/(6 \sqrt{3})$ which results in $|J_{CP}| \approx 0.0348$ for $\theta \approx 0.185$. 
A characteristic feature of the matrix in eq.(\ref{PMNSCP}), which has already been noticed in the literature \cite{mutaureflection_HS,mutaureflection_GL}, is that the absolute values 
of the entries of the second and third rows are equal, i.e. $|U_{\mu i}|=|U_{\tau i}|$
 for $i=1,2,3$. This directly leads to maximal atmospheric mixing, since $\sin \theta_{23} = \cos \theta_{23}$ and for non-vanishing $\theta_{13}$ also to a maximal Dirac phase, $\cos \delta=0$.

In the subsequent section we want to construct an explicit model based on the above choices for $Z$ and $X$, see eqs.(\ref{choiceZ}, \ref{otherX}), since this choice leads to, at least, one 
non-trivial CP phase and, simultaneously, all mixing angles are in good accordance with the experimental data for a certain value of $\theta$.
\footnote{Apart from this choice, called case I in \cite{FHZ12}, only one other choice out of the seven presented in \cite{FHZ12}, namely case IV, allows for a non-trivial Dirac phase
and a good agreement with the data on lepton mixing angles. Thus, it
 might also be interesting to implement this other case in a model.}  
Relevant aspects of the model are outlined in the following. The flavour symmetry is broken spontaneously by the
VEVs of gauge singlets. While we have chosen left-handed leptons to be in $({\bf 3'}, 1)$ under $S_4 \times Z_3$,
 the right-handed charged leptons transform as singlets: 
$e^c \sim ({\bf 1'}, 1)$, $\mu^c \sim ({\bf 1}, \omega)$ and $\tau^c \sim ({\bf 1}, \omega^2)$.
This choice turns out to be useful for accommodating the hierarchy among charged lepton masses correctly and at the same time 
does not affect the results for lepton mixing.
As we argue in section \ref{sec3}, the mass hierarchy can be naturally achieved with a judicious choice of flavons. 
The auxiliary symmetry $Z_{16}$ is invoked to sufficiently disentangle the two flavour symmetry breaking sectors and to forbid, in general, unwanted operators.
The transformation properties of the matter superfields and the MSSM Higgs doublets under $S_4 \times Z_3$ and $Z_{16}$ can be found in table \ref{tab:matter}.
\begin{table}[h!]
\begin{center}
\begin{math}
\begin{array}{|c|c|c|c|c|c|c|}
\hline
& l & e^c & \mu^c & \tau^c & h_u & h_d\\
\hline
S_4 & {\bf 3'} & {\bf 1'} & {\bf 1} & {\bf 1} & {\bf 1} & {\bf 1}\\
Z_3 & 1 & 1 & \omega & \omega^2 & 1 & 1\\
Z_{16} & 1 & \omega_8^7 & 1 & 1 & \omega_{16}^7 & 1\\
\hline
\end{array}
\end{math}
\end{center}
\begin{center}
\caption{Matter superfields and MSSM Higgs doublets $h_{u,d}$ of our model. Notice that only one of the latter is charged under the auxiliary symmetry. 
Note also that the auxiliary symmetry $Z_{16}$ is effectively a $Z_8$ symmetry at the level of operators with matter superfields, since  $(l h_u)^2$ carries a phase $\omega_8^7$.
The phases are $\omega = e^{2 \pi i/3}$, $\omega_8 =e^{2 \pi i/8}$ and  $\omega_{16} =e^{2 \pi i/16}$. \label{tab:matter}}
\end{center}
\end{table}

Since we wish to break the group $G_{CP}$ spontaneously to $G_e=Z_3^{(D)}$ and $G_\nu=Z_2 \times CP$, it is convenient to list the most general form of the VEVs that 
 flavon fields in the different representations of $S_4$ can take and which leave $G_e$ and $G_\nu$ invariant, respectively. In order to preserve
the group $G_e=Z_3^{(D)}$ flavons $\xi$ and $\xi'$ in one-dimensional representations ${\bf 1}$ and ${\bf 1'}$ of $S_4$ which have a non-vanishing VEV should transform trivially under $Z_3$.
In the case of a flavon $\chi$ being a doublet under $S_4$ the group $G_e$ can be preserved only  if $\chi$ carries a non-trivial charge under $Z_3$. The allowed
VEVs are
\begin{equation}
\label{vevEgen1}
\chi \sim ({\bf 2}, \omega) \; : \;\; \langle \chi \rangle \propto \left( \begin{array}{c} 0 \\ 1  \end{array} \right) 
\;\;\; \mbox{and} \;\;\;
\chi \sim ({\bf 2}, \omega^2) \; : \;\; \langle \chi \rangle \propto \left( \begin{array}{c} 1 \\ 0  \end{array} \right)  \; .
\end{equation}
Flavons $\psi \sim {\bf 3}$ and $\varphi \sim {\bf 3'}$ can always acquire a non-zero VEV, independent of their charge under $Z_3$. 
The latter charge, however, determines which of the components actually acquires a VEV: 
\begin{equation}
\label{vevEgen2}
\!\!\!\!\!
\psi \sim ({\bf 3}, 1) \, : \, \langle \psi \rangle \propto \left( \begin{array}{c} 1 \\ 0 \\  0 \end{array} \right) \, , \,
\psi \sim ({\bf 3}, \omega)\, : \, \langle \psi \rangle \propto \left( \begin{array}{c} 0 \\ 1 \\ 0 \end{array} \right)  \, , \,
\psi \sim ({\bf 3}, \omega^2) \, : \, \langle \psi \rangle \propto \left( \begin{array}{c} 0 \\ 0 \\ 1  \end{array} \right)  \, .
\end{equation}
The same is true for the structure of the VEVs of $\varphi \sim {\bf 3'}$. 

\noindent The VEVs that preserve $G_\nu$ are of the form
\begin{equation} 
\label{vgnu}
\langle \xi \rangle=v_\xi \; , \;\; \langle \xi' \rangle = v_{\xi'}\; , \;\;
\langle \chi \rangle = \left( \begin{array}{c} v_{\chi} \\ v^*_{\chi} \end{array} \right) \, , \;\;\; 
 \langle \psi \rangle = v_{\psi} \, \left( \begin{array}{c} 1 \\ 1 \\ 1 \end{array} \right) \, ,  \;\;
\langle \varphi \rangle = v_{\varphi} \, \left( \begin{array}{c} 1 \\ 1 \\ 1 \end{array} \right) \, , 
\end{equation}
with the parameters $v_\xi$ and $v_\varphi$ being real, $v_{\xi'}$ and $v_{\psi}$ being imaginary, while
$v_\chi$ is in general complex. Because of $G_\nu=Z_2 \times CP$ it is irrelevant for the vacuum structure preserving $G_\nu$ how
these fields transform under $Z_3$. Since left-handed leptons $l$ are in the triplet ${\bf 3'}$ of $S_4$,
only the flavons $\xi$, $\chi$ and $\varphi$ can couple to the Weinberg operator at LO, see appendix \ref{appA}. 
Upon breaking to $G_\nu$ the most general form that the neutrino mass matrix can take is
\begin{equation}
\label{mnugen}
m_\nu = \left( \begin{array}{ccc}
y_\xi v_\xi + 2 \, y_\varphi v_\varphi & y_\chi v_\chi - y_\varphi v_\varphi & y_\chi v^*_\chi - y_\varphi v_\varphi \\
y_\chi v_\chi - y_\varphi v_\varphi   &  y_\chi v^*_\chi + 2 \, y_\varphi v_\varphi & y_\xi v_\xi - y_\varphi v_\varphi\\
y_\chi v^*_\chi - y_\varphi v_\varphi & y_\xi v_\xi - y_\varphi v_\varphi & y_\chi v_\chi + 2 \, y_\varphi v_\varphi
\end{array}
\right) \, \frac{\langle h_u \rangle^2}{\La^2}
\end{equation}
with $y_\xi$, $y_\chi$ and $y_\varphi$ indicating the Yukawa couplings to the flavons $\xi$, $\chi$ and $\varphi$, respectively. 
These couplings are real because of the invariance of the original theory under a $CP$ symmetry generated by $X$.  
$\La$ is the generic high-energy cutoff scale and it suppresses all higher-dimensional operators appropriately. 
In our model it is inversely proportional to the neutrino mass scale.
As expected, the matrix in eq.(\ref{mnugen}) is diagonalized by the PMNS matrix in eq.(\ref{PMNSCP}) (remember that the charged lepton mass matrix is diagonal).
If the imaginary part of $v_\chi$ vanishes, $m_\nu$ in eq.(\ref{mnugen}) becomes
$\mu\tau$ symmetric and $G_\nu$ is promoted to $Z_2 \times Z_2 \times CP$.\footnote{As can be checked, also the second $Z_2$ group commutes with the $CP$ symmetry.} 
Thus, its relative size with respect to the other VEVs is directly related to the size of the 
 reactor mixing angle $\theta_{13}$. 
 In our model this imaginary part (and consequently $\theta_{13}$) is appropriately suppressed, since it arises effectively from an operator with two flavons, whereas 
flavons transforming as ${\bf 1}$ and ${\bf 3'}$ contribute through operators with one flavon to the matrix $m_\nu$.  
In addition, the contribution corresponding to the real part of $v_\chi$ is doubly suppressed so that
 neutrino masses depend effectively on only two parameters at LO in our model 
  (which are determined by the measured mass squared differences). As a consequence, they have to follow a normal ordering.

\section{Model with $S_4$ and $CP$}
\label{sec3}

As anticipated, we adopt the MSSM framework with left-handed lepton and right-handed charged lepton superfields and
 neglect all SUSY breaking effects in the following. The flavour symmetry is $G_f=S_4\times Z_3$ and the $CP$ transformation $X$ is defined in eq.(\ref{otherX}) for 
each representation $({\bf r}, q)$ of $G_f$. 
The transformation properties under $G_f\times Z_{16}$ of leptons and MSSM Higgs doublets 
 are collected in table 1, while those of  flavons and driving fields are presented in tables 2 and 3, respectively.
 We discuss the results for lepton masses and mixing at LO
in subsection \ref{sec31} and at NLO in subsection \ref{sec32}. 
All operators directly contributing to the lepton mass matrices are non-renormalizable in our model and are suppressed by
appropriate powers of the cutoff scale $\La$. We consider two different sets of flavons, one responsible for the breaking to $G_e$ and one for the breaking to $G_\nu$.
At LO the fields whose VEVs leave $G_e$ invariant only contribute to charged lepton masses, while those with VEVs
leaving $G_\nu$ intact generate neutrino masses. At the subleading level this changes and all flavons contribute to charged lepton as well as neutrino mass matrices.

\subsection{Leading order results}
\label{sec31}

 As outlined in section \ref{sec2}, left-handed leptons form a triplet ${\bf 3'}$, while right-handed charged 
leptons are singlets under $S_4 \times Z_3$. 
With this assignment the different charged lepton masses can arise at the one-, two- and higher-flavon level in our model and thus 
are naturally hierarchical. As flavon fields $\varphi_E \sim ({\bf 3'}, \omega)$ and $\chi_E \sim ({\bf 2}, \omega)$ under $S_4\times Z_3$ are included.\footnote{Adding $\chi_E$ is mainly 
necessary in order to achieve the correct vacuum alignment, see section \ref{sec4}.}
The most relevant operators for charged lepton masses are then\footnote{We indicate the contraction to a trivial singlet ${\bf 1}$ of $S_4$ by $(\cdots)$.}
\begin{equation}
\label{weLO}
w_e = y _\tau (l \varphi_E) \tau^c h_d/\La + y_{\mu, 1} (l \varphi_E^2) \mu^c h_d/\La^2 + y_{\mu, 2} (l \chi_E \varphi_E) \mu^c h_d/\La^2 \; .
\end{equation}
Since we impose $CP$ as symmetry on the theory in the unbroken phase, all couplings in $w_e$ and in the following superpotentials are real.
According to the results of section \ref{sec4}, the flavons $\chi_E$ and $\varphi_E$ acquire at LO the VEVs 
\begin{equation}
\label{vevEmodel}
\langle \chi_E \rangle = v_{\chi_E} \, \left( \begin{array}{c} 0  \\ 1 \end{array} \right) \;\;\; \mbox{and} \;\;\;
\langle \varphi_E \rangle = v_{\varphi_E} \, \left( \begin{array}{c} 0  \\ 1 \\ 0 \end{array} \right)
\end{equation}
with $v_{\chi_E}$ and $v_{\varphi_E}$ being complex, see eq.(\ref{vErels}). If we compare the form of the VEVs in eq.(\ref{vevEmodel}) with the VEVs given in eqs.(\ref{vevEgen1}, \ref{vevEgen2}), we see that
$\langle \chi_E \rangle$ and $\langle \varphi_E \rangle$ indeed leave the subgroup $G_e=Z_3^{(D)}$ of $S_4 \times Z_3$ invariant. Plugging them into the operators in eq.(\ref{weLO}), 
we find $m_l$ to be diagonal
in flavour space and to only have non-zero entries in the (22) and (33) elements that correspond to the muon and tau lepton masses
\begin{equation}
\label{mchLO}
m_\mu =\left| (2~y_{\mu,1}  v_{\varphi_E} + y_{\mu,2}  v_{\chi_E}) \,  \frac{v_{\varphi_E}}{\La^2} \right| \, \langle h_d \rangle \;\;\; \mbox{and} \;\;\;
m_\tau = \left| y_\tau \frac{v_{\varphi_E}}{\La} \right| \, \langle h_d \rangle \; ,
\end{equation}
respectively. Their relative hierarchy is correctly reproduced for 
\begin{equation}
\label{ve}
|v_{\chi_E}|,  |v_{\varphi_E}| \sim \lambda^2 \, \La \;\;\; \mbox{with} \;\;\; \lambda \approx 0.2 \; .
\end{equation}  
Since $|v_{\varphi_E}|/\La \approx \lambda^2 \approx 0.04$, $\tan\beta= \langle h_u \rangle/ \langle h_d \rangle$ is expected to lie in the interval $2 \lesssim \tan\beta \lesssim 12$ (for $0.5 \lesssim y_\tau \lesssim 3$).
The electron remains massless at this level and its mass is generated via higher-dimensional operators, as discussed in subsection \ref{sec32}.

\begin{table}[h!]
\begin{center}
\begin{math}
\begin{array}{|c|c|c|c|c|c|c|}
\hline
\rule[0.17in]{0cm}{0cm} & \chi_E & \varphi_E & \xi_N, \tilde{\xi}_N & \chi_N & \varphi_N & \xi'_N\\
\hline
S_4 & {\bf 2} & {\bf 3'} & {\bf 1} & {\bf 2} & {\bf 3'} & {\bf 1'}\\
Z_3 & \omega & \omega &1 & 1 & 1 & 1\\
Z_{16} & 1 & 1 & \omega_8 & \omega_8^5 & \omega_8 & \omega_8^4\\
\hline
\end{array}
\end{math}
\end{center}
\begin{center}
\caption{Flavons and their transformation properties under $G_f\times Z_{16}$. The flavons labelled with $E$ are charged under $Z_3$ and neutral under $Z_{16}$, whereas those with an index $N$
only carry a non-trivial charge under $Z_{16}$. The phases are $\omega= e^{2 \pi i/3}$ and $\omega_8 = e^{2 \pi i/8}$.
\label{tab:flavons}}
\end{center}
\end{table}

As mentioned, neutrino masses arise in our model from the Weinberg operator. The flavon multiplets $\xi_N$, $\tilde{\xi}_N$ and $\varphi_{N}$ are in 
 $({\bf 1},1)$ and in $({\bf 3'},1)$ under $S_4 \times Z_3$, respectively, and each of them can be coupled independently to neutrinos, while the fields $\xi'_N \sim ({\bf 1'}, 1)$ and $\chi_{N} \sim ({\bf 2},1)$ 
 couple to the latter in the combination $\chi_N \xi'_N$. This situation can be arranged by an appropriate choice of the charges under the auxiliary symmetry $Z_{16}$, 
 see table \ref{tab:flavons}. Note that the two fields $\xi_N$ and $\tilde{\xi}_N$ have the same quantum numbers and both are necessary in order to achieve the correct vacuum alignment with the potential discussed
 in section \ref{sec4}. However, only one of them acquires a non-vanishing VEV, see eq.(\ref{nuvevs}). The lowest dimensional operators responsible for neutrino masses are
 \begin{equation}
 \label{wnuLO}
 w_\nu = y_{\nu,1} (l l) \xi_N h_u^2/\La^2 +  \tilde{y}_{\nu,1} (l l)  \tilde{\xi}_N h_u^2/\La^2 +y_{\nu,2} (l l \varphi_N) h_u^2/\La^2 + y_{\nu,3} (l l \chi_N \xi'_N) h_u^2/\La^3 \; .
 \end{equation}
 As shown in section \ref{sec4}, the flavons $\xi_N$, $\tilde{\xi}_N$, $\xi'_N$, $\chi_N$ and $\varphi_{N}$ develop the following VEVs at LO
 \begin{equation} 
\langle \tilde{\xi}_N \rangle =0~ , \;\;\; 
\langle \xi_N \rangle = v_{\xi_N} ~ , \;\;\;  
\langle \xi'_N \rangle =  v_{\xi'_N} ~ , \;\;\;
\langle \chi_N \rangle = v_{\chi_N} \, \left( \begin{array}{c} 1 \\ 1 \end{array} \right) ~, \;\;\;
\langle \varphi_N \rangle = v_{\varphi_N} \,  \left( \begin{array}{c} 1 \\ 1 \\ 1 \end{array} \right)
\label{nuvevs}
\end{equation}
with $v_{\xi_N}$, $v_{\chi_N}$ and $v_{\varphi_N}$ being complex, while $v_{\xi'_N}$ is imaginary.
The VEVs $v_{\xi_N}$, $v_{\chi_N}$ and $v_{\varphi_N}$ are related via the parameters of the flavon superpotential, see eq.(\ref{vNrels}),
and thus, under very mild assumptions on these parameters, they carry the same phase (up to $\pi$). 
The neutrino mass matrix $m_\nu$ takes the form
\begin{equation}
\label{mnuLO}
m_\nu = \left( \begin{array}{ccc}
t_\nu + 2 \, u_\nu & - u_\nu - i x_\nu& -u_\nu + i x_\nu\\
- u_\nu - i x_\nu & 2 \, u_\nu + i x_\nu& t_\nu - u_\nu\\
- u_\nu + i x_\nu & t_\nu - u_\nu & 2 \, u_\nu - i x_\nu
\end{array}
\right) \, \frac{\langle h_u \rangle^2}{\La}
\end{equation}
with  
\begin{equation}
\label{paratux}
t_\nu = y_{\nu,1} \, \frac{v_{\xi_N}}{\La}  \; , \;\; u_\nu = y_{\nu,2} \, \frac{v_{\varphi_N}}{\La}  \; , \;\; x_\nu =  i \, y_{\nu, 3} \, \frac{v_{\xi'_N} v_{\chi_N}}{\La^2} \; .
\end{equation}
The common phase of $v_{\xi_N}$, $v_{\chi_N}$ and $v_{\varphi_N}$ is unphysical as regards lepton masses and mixing at this level,
since it can be factored out in $m_\nu$ in eq.(\ref{mnuLO}). Thus, the parameters
$t_\nu$, $u_\nu$ and $x_\nu$ can be considered as real. Among the leading contributions coming from the operators in eq.(\ref{wnuLO}) the one
comprising the fields $\xi'_N$ and $\chi_N$ renders $m_\nu$ invariant only under $G_\nu=Z_2 \times CP$ and not
$Z_2 \times Z_2 \times CP$.
Note that the matrix in eq. (\ref{mnuLO}) slightly differs from the one in eq. (\ref{mnugen}), since, after extracting the overall phase, the combination 
 $\chi_N \xi'_N$ only develops an imaginary VEV. The latter 
  is responsible for a non-vanishing value of $\theta_{13}$ and is naturally smaller than the other contributions to $m_\nu$,
 because it arises from an operator with two flavons. The correct size of $\theta_{13}$ is achieved, if the relative strength of this contribution
is suppressed by a factor $\lambda$ with respect to the ones from the first three operators in eq.(\ref{wnuLO}). This is the case, if
 we choose\footnote{Choosing all flavon VEVs associated with the neutrino sector to be of order $\lambda \, \Lambda$ is only convenient, but a priori not necessary.}
 \begin{equation}
 \label{vnu}
|v_{\xi_N}|, |v_{\xi'_N}|, |v_{\chi_N}|,  |v_{\varphi_N}| \sim \lambda \, \La \; .
\end{equation}

As mentioned in section \ref{sec2}, the neutrino mass matrix $m_\nu$ is diagonalized by a mixing matrix of the form given in eq.(\ref{PMNSCP}).
At LO this matrix gives the PMNS mixing matrix, since the charged lepton mass matrix is diagonal at this level. Thus, maximal atmospheric mixing and a maximal Dirac phase are obtained.
 The angle $\theta$ is directly related to $\theta_{13}$, see eq.(\ref{mixingangles}), and it depends on the two parameters $u_\nu$ and $x_\nu$ in our model
\begin{equation}
\tan 2 \theta = \frac{x_\nu}{\sqrt{3} \, u_\nu} \; .
\end{equation}
Our predictions for the mixing parameters, especially maximal atmospheric mixing and the maximal Dirac phase, can be tested in various ways: indirectly, with global fits of all neutrino data, e.g. \cite{global_latest}, as well as 
directly, with the experiments T2K and NO$\nu$A which could determine after five to six years of data taking the octant of the atmospheric mixing angle at  the $(2 \div 3) \, \sigma$ 
level \cite{T2KNOvA}, depending on the actual value of $\theta_{23}$. The same experiments can possibly exclude certain ranges of the Dirac phase
within the next ten years, depending on $\theta_{23}$ and the neutrino mass hierarchy \cite{T2KNOvA_dCP}. Indeed, T2K has recently reported a slight preference for $\delta$ close to $3 \, \pi/2$
 \cite{T2Kdelta}, when their data are combined with the measurements of the reactor experiments. This preference 
  is compatible with the $1 \, \sigma$ preferred range for $\delta$, $\pi \lesssim \delta \lesssim 1.9 \, \pi$, found in global fits \cite{global_latest,global_Bari}. 
Dedicated experiments like LBNE \cite{LBNE} and Hyper-Kamiokande \cite{H_K} will allow for a measurement of the Dirac phase with a certain precision and have thus the potential to rule out our model.
The neutrino masses read
\begin{equation} \label{numass_LO}
m_1 = \left\vert t_\nu + \tilde{u}_\nu \right\vert \,  \frac{\langle h_u \rangle^2}{\La} \; , \;\;
m_2 = \left\vert t_\nu \right\vert \, \frac{\langle h_u \rangle^2}{\La}  \; , \;\;
m_3 = \left\vert t_\nu - \tilde{u}_\nu \right\vert \,  \frac{\langle h_u \rangle^2}{\La}
\end{equation}
and we have defined
\begin{equation}
\tilde{u}_\nu = \sqrt{3}  \, \mbox{sign}(u_\nu \cos 2 \theta) \, \sqrt{3 u_\nu^2 + x_\nu^2} \; .
\end{equation}
Neutrino masses as in eq.(\ref{numass_LO}) imply normal mass hierarchy: the requirement $m_2 > m_1$ entails $t_\nu \tilde u_\nu <0$ and $2 \, |t_\nu| \gtrsim |\tilde{u}_\nu|$ and, consequently, 
the atmospheric mass squared difference
\begin{equation}
\label{dm2atm_LO}
\Delta m_{\mathrm{atm}}^2 = m_3^2 - m_1^2 = - 4 \, t_\nu \tilde{u}_\nu \, \left( \frac{\langle h_u \rangle^2}{\La} \right)^2
\end{equation}
is always positive. The smallness of the ratio $\Delta m_{\mathrm{sol}}^2/\Delta m_{\mathrm{atm}}^2$ is achieved by an appropriate tuning of $2 \, t_\nu + \tilde{u}_\nu$.
The normal mass hierarchy is an important consequence of the fact that $\chi_N \xi'_N$ only develops an imaginary VEV, which implies that $m_\nu$ contains only three real parameters at LO, see
eq. (\ref{mnuLO}).

Since the neutrino masses only depend on the two real parameters $t_\nu$ and $\tilde{u}_\nu$, both of them can be fixed by the measured values of the mass squared differences and 
the absolute scale of neutrino masses and also their sum are predicted. Taking the best fit values for $\Delta m_{\rm sol}^2$ and  $\Delta m_{\rm atm}^2$ quoted in \cite{global_latest}
we get 
\begin{equation}
m_1 \approx 0.016 \, \mathrm{eV} \;\; , \;\;\; m_2 \approx 0.018 \, \mathrm{eV} \;\; , \;\;\; m_3 \approx 0.052 \, \mathrm{eV}
\end{equation}
and 
\begin{equation}
\Sigma \, m_\nu \approx 0.086 \, \mathrm{eV} \; .
\end{equation}
This value of the sum of the neutrino masses is well compatible with the latest results from the Planck satellite \cite{Planck}. For $\tan\beta \approx 10$
and $\La \approx 3 \times 10^{14} \, \mathrm{GeV}$, $\langle h_u \rangle^2/\La = 0.1 \, \mathrm{eV}$ follows and the parameters $t_\nu$, $u_\nu$ and $x_\nu$ are of order
$\lambda$, $\lambda$ and $\lambda^2$, respectively. The quantity $m_\beta$ which is measured in $\beta$ decay experiments, is $m_\beta \approx 0.018 \, \mathrm{eV}$  
which is an order of magnitude below the expected sensitivity of the KATRIN experiment \cite{Katrin}. Thus, a positive signal of KATRIN or of the proposed experiments Project 8,  ECHo, MARE and PTOLEMY, could 
rule out our model, see e.g. \cite{mbeta_MH}. Also an experiment like EUCLID \cite{euclid} which could reach a $1 \, \sigma$ precision 
of  0.01 eV for the sum of the neutrino masses \cite{euclid_reach} could disfavour our model. An exclusion of the wrong mass ordering is expected at the $3 \, \sigma$ level with experiments like 
PINGU, ORCA, JUNO and RENO50 (and LBNE), within the next ten to fifteen years, for recent studies see \cite{MH_june_nov2013}. At any rate the
 information on the
mass hierarchy will be useful for improving the reach of measuring e.g. CP violation.

Both Majorana phases $\alpha$ and $\beta$ are determined by the form of the neutrino masses in eq.(\ref{numass_LO}) and by the requirement $t_\nu \tilde u_\nu < 0$
\begin{equation}
\alpha= \pi \,\,\, , \,\, \beta=\pi 
\end{equation}
and the quantity $m_{ee}$ which encodes neutrinoless double $\beta$ decay \cite{0nubb_review}  is  $m_{ee} \approx 0.003 \, \mathrm{eV}$. 
Such a low value is experimentally very challenging \cite{0nubb_review,EXO,KamLandZen,GERDA}. On the other hand, if a positive signal is found, our model would be disfavoured.

\subsection{Next-to-leading order results}
\label{sec32}

In this subsection we address relevant contributions arising from shifts in the LO vacuum and from operators with more flavons than those discussed above.
It is important to keep in mind in the following that we have chosen different expansion parameters in the charged lepton and neutrino sectors,
see eqs. (\ref{ve}, \ref{vnu}).

We first discuss the most important corrections to the neutrino mass matrix. These arise from the shift in the VEV of the flavon $\chi_N$, see eq.(\ref{VEVchiNNLO}), if
plugged into the fourth operator of eq.(\ref{wnuLO}), and lead to a contribution to the neutrino mass matrix in eq.(\ref{mnuLO}) which is suppressed by $\lambda^2$ with respect to those coming from the
first three terms in eq.(\ref{wnuLO}). This new contribution cannot be absorbed via a re-definition of the three parameters $t_\nu$, $u_\nu$ and $x_\nu$ in $m_\nu$, but instead
requires a fourth parameter $p_\nu$. This parameter is real, since the corrections to the VEV of $\chi_N$ as well as $v_{\xi'_N}$ are both imaginary and all 
couplings are real. Furthermore, the contribution is $\mu\tau$-symmetric in flavour space.
The corrected neutrino mass matrix thus reads
\begin{equation}
\label{mnuNLO}
m_\nu^{\mathrm{NLO}}=\left( \begin{array}{ccc}
t_\nu + 2 u_\nu & - u_\nu - i x_\nu + p_\nu& -u_\nu + i x_\nu + p_\nu\\
- u_\nu - i x_\nu + p_\nu& 2 u_\nu + i x_\nu + p_\nu& t_\nu - u_\nu\\
- u_\nu + i x_\nu + p_\nu& t_\nu - u_\nu & 2 u_\nu - i x_\nu + p_\nu
\end{array}
\right) \, \frac{\langle h_u \rangle^2}{\La}
\end{equation}
with 
\begin{equation}
\label{parapnu}
p_\nu =  i \,  y_{\nu, 3} \, \alpha \, \frac{v_{\xi'_N} v_{\chi_N}}{\La^2}  \, \lambda \, 
\end{equation}
being real, since $i \, v_{\xi'_N}$ is real. This matrix has the most general form which is compatible with the preservation of the group $G_\nu$ in the neutrino sector (modulo the overall phase which also shows up in the parameter $p_\nu$
and thus can be factored out). Consequently, the results for the mixing parameters remain unchanged, whereas the neutrino 
masses, see eq.(\ref{numass_LO}), acquire corrections proportional to $p_\nu$.
Since $p_\nu$ is suppressed by $\lambda^2$ with
respect to $t_\nu$ and $u_\nu$ (and thus $\tilde u_\nu$), the correction to the LO results for the neutrino masses is of relative order $\lambda^2$.

The largest correction to the mass matrix $m_\nu$ which alters its structure in such a way that it is no longer compatible with the preservation of the residual symmetry $G_\nu$
arises at order $\lambda^6$ in units of $\langle h_u \rangle^2/\La$.\footnote{Here and in the following we assume 
that the LO vacuum of the flavons $\xi_N$, $\tilde\xi_N$, $\chi_N$, $\varphi_N$ and $\xi'_N$ actually preserves $G_\nu$, i.e. in our case that the
VEVs $v_{\xi_N}$, $v_{\chi_N}$ and $v_{\varphi_N}$ are real.} It originates from shifting the vacuum of the field $\xi'_N$ so that $\langle \xi'_N \rangle$ gets a small real part, being
suppressed by $\lambda^4$ with respect to $v_{\xi'_N}$ (which is imaginary). This shift contributes to $m_\nu$ through the fourth operator in eq.(\ref{wnuLO}) and leads to unequal real parts of the (12) and (13) and (22) and (33) elements of $m_\nu$,
respectively, whose difference is of relative order $\lambda^5$ compared to the LO term in these elements.
If we include operators with more flavons and compute their contribution using the LO vacuum of these fields, corrections 
to $m_\nu$ that break  the residual symmetry $G_\nu$ of $m_\nu^{\mathrm{NLO}}$ in eq.(\ref{mnuNLO}) are generated
at order $\lambda^7$ (and smaller) in units of $\langle h_u \rangle^2/\La$. The most relevant ones contain 
three flavons belonging to the set $\{ \chi_E, \varphi_E \}$ and one being either $\xi_N$, $\tilde{\xi}_N$ or $\varphi_N$.
Both types of corrections will affect our results for the mixing parameters. However, they are suppressed by at least $\lambda^4$ with respect to the parameters in $m_\nu$ in eq.(\ref{mnuLO})
 and so can be safely neglected.

In the same fashion we can analyze the subleading contributions to the charged lepton mass matrix which is diagonal at LO and does not allow for a non-zero electron mass. The latter instead is generated in two ways: first through
operators with five flavons with an index $N$, if we take into account the shift in the VEV of the flavon $\chi_N$ (in the following we only mention operators that actually give a non-zero contribution to the electron mass, omit all (real) Yukawa couplings
and do not specify whether these operators arise from only one independent $S_4$ contraction or whether there are several independent ones)
\begin{equation} 
l e^c  \xi_N^3 \chi_N \varphi_N h_d/\La^5 + l e^c \xi_N \chi_N \varphi_N^3 h_d/\La^5  + l e^c \xi_N \chi_N^3 \varphi_N h_d/\La^5 
\end{equation}
and secondly through operators with six flavons with an index $N$ (one of them being $\xi'_N$, an even number of fields $\chi_N$ and the rest either $\xi_N$, $\tilde{\xi}_N$ or $\varphi_N$), if we use the LO vacuum of these fields,
\begin{eqnarray}
\nonumber 
&&l e^c \xi_N^4 \xi'_N \varphi_N h_d/\La^6 + l e^c \xi_N^2 \xi'_N \varphi_N^3 h_d/\La^6 + l e^c \xi'_N \varphi_N^5 h_d/\La^6 
\\
&+& l e^c \xi_N^2 \xi'_N \chi_N^2 \varphi_N h_d/\La^6 + l e^c  \xi'_N \chi_N^2 \varphi_N^3 h_d/\La^6 + l e^c \xi'_N \chi_N^4 \varphi_N h_d/\La^6 \; .
\end{eqnarray}
All these operators induce contributions to the $(1i)$ elements, $i=1,2,3$ of the charged lepton mass matrix that are of the order $\lambda^6 \, \langle h_d \rangle$ and equal.
They lead to an electron mass of order $\lambda^6 \langle h_d \rangle$ which is in agreement with the observed mass hierarchy among the charged leptons. 
The reason for these operators to give rise to a contribution that is equal for all three elements of the first row of $m_l$ is the following:  $l e^c h_d \sim ({\bf 3}, 1)$ under $S_4 \times Z_3$ and 
 the most general form of the VEV of a field in the representation ${\bf 3}$ of $S_4$ which leaves the group $G_\nu$ invariant is a vector with three equal entries, see eq.(\ref{vgnu}), so that the contribution to the
 charged lepton mass matrix involving only fields with an index $N$ leads to equal $(1i)$ elements.

Also the other elements of the charged lepton mass matrix are corrected at a subleading level. One type of corrections originates from the shifts in the VEVs of the fields $\chi_E$ and $\varphi_E$, see eq.(\ref{VEVsENLO}), 
and results in $(31)$ and $(32)$ elements of order $\lambda^6$ and $(21)$ and $(23)$ elements of order $\lambda^8$ in units of $\langle h_d \rangle$. A second source of corrections are operators with more flavons which turn out
to have the following structure: an operator with $\mu^c$ comprises two fields from the set $\{ \chi_E, \varphi_E \}$ in order to saturate the $Z_3$ charge and either one field $\chi_N$ together with three flavons of the type 
$\{ \xi_N, \tilde{\xi}_N, \varphi_N \}$ or three fields $\chi_N$ and one field $\xi_N$, $\tilde{\xi}_N$ or $\varphi_N$, whereas an operator with $\tau^c$ only requires one field from the set $\{ \chi_E, \varphi_E\}$, while otherwise maintaining the
same form as the operator with $\mu^c$, e.g.
\begin{equation}
l \mu^c \chi_E^2 \xi_N^2 \varphi_N \chi_N h_d/\La^6 + l \tau^c \varphi_E \xi_N^3 \chi_N h_d/\La^5 \; .
\end{equation}
Taking into account the LO form of the flavon VEVs, we see that such operators contribute at the same level to the charged lepton mass matrix as the leading operators, if the latter are computed with the shifted vacuum of $\chi_E$ 
and $\varphi_E$. Eventually we find the order of magnitudes of the different entries of the charged lepton mass matrix, including subleading corrections, to be
\begin{equation}
m_l^{\mathrm{NLO}} \sim \left( \begin{array}{ccc}
\lambda^6 & \lambda^6 & \lambda^6\\
\lambda^8 & \lambda^4 & \lambda^8\\
\lambda^6 & \lambda^6 & \lambda^2
\end{array}
\right) \; \langle h_d \rangle \; .
\end{equation}
The charged lepton masses thus follow the correct hierarchy
\begin{equation}
m_e : m_\mu : m_\tau \approx \lambda^4 : \lambda^2 : 1
\end{equation}
with $m_\tau \approx \lambda^2 \, \langle h_d \rangle$ and we can estimate the size of the contribution to the lepton mixing angles coming from the charged lepton sector as
\begin{equation}
\theta_{ij}^l \sim \lambda^4 \;\;\; \mbox{with} \;\;\; ij=12,13,23 \; .
\end{equation}
Such corrections have a small impact on our results for the lepton mixing parameters and can be neglected.

\section{Flavon superpotential}
\label{sec4}

In order to construct the superpotential responsible for the alignment of the flavon VEVs we assume the existence of an $R$-symmetry $U(1)_R$ \cite{AF2}
under which matter superfields carry charge +1, fields acquiring a non-vanishing vacuum (i.e. $h_{u,d}$ and all flavons) are uncharged and so-called driving fields,
indicated with the superscript ``$0$" carry charge +2. In this way all terms in the superpotential either contain two matter superfields (like the terms leading to lepton masses)
or one driving field, like those to be discussed in this section. In the limit of unbroken SUSY, the F-terms of these fields have to vanish and in this way the vacuum
of the flavons gets aligned.\footnote{The vanishing of the F-terms of the flavons can be achieved for vanishing VEVs of the driving fields.}
 The driving fields necessary for our construction of the flavon superpotential are given in table \ref{tab:driving}. Their transformation properties under
the flavour and auxiliary symmetries are chosen in such a way that, at the renormalizable level, the vacua of the sets of fields $\{ \chi_E , \varphi_E \}$, $\{ \xi_N, \tilde{\xi}_N, \chi_N, \varphi_N \}$
and $\xi'_N$ are aligned separately: the vacuum of the first set of fields spontaneously breaks $G_{CP}$ to $Z_3^{(D)}$, while the one of the second set breaks $G_{CP}$ to $Z_2 \times G_\nu$
(up to the overall phase of the VEVs, see eq.(\ref{vNrels}), which is irrelevant for lepton masses and mixing at LO) that is broken to $G_\nu$, if the vacuum of $\xi'_N$ is also considered.

\begin{table}[h!]
\begin{center}
\begin{math}
\begin{array}{|c|c|c|c|c|c|c|}
\hline
\rule[0.17in]{0cm}{0cm}  & \xi^0_E \, , \, \tilde{\xi}^0_E & \chi^0_E & \xi^0_N & \chi^0_N & \varphi^0_N & \tilde{\xi}^0_N\\
\hline
S_4 & {\bf 1} & {\bf 2} & {\bf 1} & {\bf 2} & {\bf 3'} & {\bf 1}\\
Z_3 & \omega & \omega & 1 & 1 & 1 & 1\\
Z_{16} & 1 & 1 & \omega_4^3 & \omega_4^3 & \omega_4^3 & 1\\
\hline
\end{array}
\end{math}
\end{center}
\begin{center}
\caption{Driving fields and their transformation properties under $G_f\times Z_{16}$. The fields labelled by $E$ are charged under $Z_3$ and neutral under $Z_{16}$. The opposite holds for fields labelled by $N$. The phases are  $\omega= e^{2 \pi i/3}$ and $\omega_4 = e^{2 \pi i/4}$. \label{tab:driving}}
\end{center}
\end{table}
%

\subsection{Leading order results}
\label{sec41}

We can divide the flavon superpotential $w_{fl}$ in three parts $w_{fl,e}$, $w_{fl,\nu}$ and $w_{fl,\xi}$
\begin{equation}
\label{wfl_LO}
w_{fl}= w_{fl,e} + w_{fl,\nu} + w_{fl,\xi}
\end{equation}
with
\begin{equation}
\label{wfle_LO}
w_{fl,e} = a_e \, \xi^0_E (\chi_E \chi_E) + \tilde{a}_e \, \tilde{\xi}^0_E (\varphi_E \varphi_E) + b_e \, (\chi^0_E \chi_E \chi_E) + c_e \, (\chi^0_E \varphi_E \varphi_E) \; ,
\end{equation}
\begin{eqnarray}
w_{fl, \nu} &=& a_\nu \, \xi^0_N \xi^2_N + \tilde{a}_\nu \, \xi^0_N \tilde{\xi}^2_N + \bar{a}_\nu  \, \xi^0_N \xi_N \tilde{\xi}_N + b_\nu \, \xi^0_N (\chi_N \chi_N)  + c_\nu \, \xi^0_N (\varphi_N \varphi_N) 
 \nonumber
\\ \label{wflnu_LO}
&&+ d_\nu \, (\chi^0_N \chi_N \chi_N) + e_\nu \, (\chi^0_N \varphi_N \varphi_N) +  f_\nu \, \tilde{\xi}_N (\varphi_N^0 \varphi_N) + g_\nu \, (\varphi_N^0 \varphi_N \varphi_N)
\end{eqnarray}
and
\begin{equation}
\label{wflxi_LO}
w_{fl,\xi} = \tilde{\xi}^0_N M^2 + a_\xi \tilde{\xi}^0_N (\xi'_N \xi'_N) \; .
\end{equation}
A few things are noteworthy: all couplings are real, since the theory is invariant under $CP$ in the unbroken phase;
the driving fields $\xi^0_E$ and $\tilde{\xi}^0_E$ are defined in such a way that the former only couples to $(\chi_E \chi_E)$ and the latter only to $(\varphi_E \varphi_E)$;
similarly, we use the freedom to define the fields $\xi_N$ and $\tilde{\xi}_N$ in such  a way that only $\tilde{\xi}_N$ couples to $\varphi_N$ (at the renormalizable level); moreover, 
we neglect all terms containing the MSSM Higgs doublets $h_{u,d}$, since these are completely irrelevant in the discussion of the
alignment of the flavon VEVs which are typically of order $10^{13}$ GeV.  

The F-terms of the driving fields with an index $E$ are of the form
\begin{subequations}
\label{Ftermse}
\begin{eqnarray}
\frac{\partial w_{fl}}{\partial \xi^0_E} &=& 2 a_e \, \chi_{E,1} \chi_{E,2} \; ,
\\ 
\frac{\partial w_{fl}}{\partial \tilde{\xi}^0_E} &=& \tilde{a}_e \, (\varphi^2_{E,1} + 2 \varphi_{E,2} \varphi_{E,3}) \; ,
\\ 
\frac{\partial w_{fl}}{\partial \chi^0_{E,1}} &=& b_e \, \chi^2_{E,1} + c_e \, (\varphi^2_{E,3} + 2 \varphi_{E,1} \varphi_{E,2})  \; ,
\\ 
\frac{\partial w_{fl}}{\partial \chi^0_{E,2}} &=& b_e \, \chi^2_{E,2} + c_e \, (\varphi^2_{E,2} + 2 \varphi_{E,1} \varphi_{E,3})  \; .
\end{eqnarray}
\end{subequations}
Equating these to zero and solving for the VEVs of the flavons $\chi_E$ and $\varphi_E$ we get
\begin{equation}
\label{vELO}
\langle \chi_E \rangle = v_{\chi_E} \, \left( \begin{array}{c} 0  \\ 1 \end{array} \right) \;\;\; \mbox{and} \;\;\;
\langle \varphi_E \rangle = v_{\varphi_E} \, \left( \begin{array}{c} 0  \\ 1 \\ 0 \end{array} \right)
\end{equation}
up to symmetry transformations belonging to $S_4$. Obviously, also the trivial vacuum is a solution which, however, can be easily excluded by requiring that at least one of the flavons $\chi_E$ and $\varphi_E$ acquires a non-vanishing VEV. 
The two VEVs $v_{\chi_E}$ and $v_{\varphi_E}$ are related through
\begin{equation}
\label{vErels}
v_{\varphi_E}^2 = - \frac{b_e}{c_e} \, v_{\chi_E}^2 
\end{equation}
with $v_{\chi_E}$ being a free parameter which is in general complex. The presence of this parameter indicates a flat direction which exists in the flavon potential in the SUSY limit.
Since $v_{\chi_E}$ and $v_{\varphi_E}$ are directly related, it is natural to expect them to be of the same order of magnitude, as we did in the preceding discussion. Being their size related to a flat direction we, however, cannot fix
their absolute size, but choose this by hand in order to correctly accommodate the hierarchy among the tau and the muon mass, see eqs.(\ref{mchLO},\ref{ve}).

Analogously, the F-terms of the driving fields $\xi^0_N$, $\chi^0_N$ and $\varphi^0_N$ are given by
\begin{subequations}
\begin{eqnarray}
\label{FNa}
\frac{\partial w_{fl}}{\partial \xi^0_N} &=& a_\nu \, \xi^2_N + \tilde{a}_\nu \, \tilde{\xi}^2_N + \bar{a}_\nu  \, \xi_N \tilde{\xi}_N + 2~ b_\nu \, \chi_{N,1} \chi_{N,2}
+ c_\nu \, (\varphi_{N,1}^2 + 2 \varphi_{N,2} \varphi_{N,3})  \; ,
\\ 
\label{FNb}
\frac{\partial w_{fl}}{\partial \chi^0_{N,1}} &=& d_\nu \, \chi_{N,1}^2 + e_\nu \, (\varphi_{N,3}^2 + 2 \varphi_{N,1} \varphi_{N,2})  \; ,
\\ 
\label{FNc}
\frac{\partial w_{fl}}{\partial \chi^0_{N,2}} &=& d_\nu \, \chi_{N,2}^2 + e_\nu \, (\varphi_{N,2}^2 + 2 \varphi_{N,1} \varphi_{N,3})  \; ,
\\ 
\frac{\partial w_{fl}}{\partial \varphi^0_{N,1}} &=& f_\nu \,  \tilde{\xi}_N \varphi_{N,1} + 2 \, g_\nu \,  (\varphi_{N,1}^2  - \varphi_{N,2} \varphi_{N,3})  \; ,
\\ 
\frac{\partial w_{fl}}{\partial \varphi^0_{N,2}} &=&  f_\nu \,  \tilde{\xi}_N \varphi_{N,3} + 2 \, g_\nu \,  (\varphi_{N,2}^2  - \varphi_{N,1} \varphi_{N,3})  \; ,
\\ 
\frac{\partial w_{fl}}{\partial \varphi^0_{N,3}} &=&  f_\nu \,  \tilde{\xi}_N \varphi_{N,2} + 2 \, g_\nu \,  (\varphi_{N,3}^2  - \varphi_{N,1} \varphi_{N,2})  \; .
\end{eqnarray}
\end{subequations}
Equating these to zero, we find several possible solutions for the vacuum. These can be classified and have different features:  
one class is formed by the trivial vacuum, one class by solutions with vanishing VEVs of the fields $\chi_N$ and $\varphi_N$, a third class comprises
the vacua in which the VEV of $\chi_N$ vanishes, while the other flavons acquire non-zero VEVs, and the last class requires $\langle \tilde{\xi}_N \rangle=0$
and all other VEVs to be different from zero. We are indeed interested in this last class and, as one sees, it can be selected by requiring that (at least) the 
flavon $\chi_N$ being in the two-dimensional representation of $S_4$ acquires a non-zero VEV. Then all solutions belonging to this class are related by 
$S_4$ transformations and/or by a change of the relative sign of the VEVs of the two components of the flavon $\chi_N$. The latter feature can be traced back to the fact that the expressions in eqs.(\ref{FNb},\ref{FNc}) 
only depend quadratically on $\chi_{N,i}$. Given these different possibilities, we choose as vacuum
\begin{equation} 
\langle \xi_N \rangle =v_{\xi_N} \; , \;\;\;  \langle \chi_N \rangle = v_{\chi_N} \, \left( \begin{array}{c} 1 \\ 1 \end{array} \right) \; , \;\;\;
\langle \varphi_N \rangle = v_{\varphi_N} \,  \left( \begin{array}{c} 1 \\ 1 \\ 1 \end{array} \right)
\end{equation}
and $\langle \tilde{\xi}_N \rangle =0$.\footnote{If the vacuum with a relative sign among the VEVs of the two components of $\chi_N$ was chosen, 
 the resulting neutrino mass matrix at LO, see eq.(\ref{mnuLO}), would turn out  to be $\mu\tau$ symmetric. 
 As a consequence, the reactor mixing angle $\theta_{13}$ would vanish, in contradiction to experimental observation, and the 
 Dirac phase would become unphysical. This example shows that not all degenerate vacua are physically equivalent in our model.}
As can be derived from eqs.(\ref{FNa}-\ref{FNc}), the VEVs $v_{\xi_N}$, $v_{\chi_N}$ and $v_{\varphi_N}$ are related through
\begin{equation}
\label{vNrels}
v^2_{\xi_N} = \frac{3}{a_\nu} \, \left(2~ \frac{b_\nu e_\nu}{d_\nu} - c_\nu \right) \, v_{\varphi_N}^2
\;\;\; \mbox{and} \;\;\;
  v_{\chi_N}^2= - 3 \, \frac{e_\nu}{d_\nu} \, v_{\varphi_N}^2
\end{equation}
with $v_{\varphi_N}$ parametrizing the second flat direction of our flavon potential in the SUSY limit. 
This parameter is in general complex so that also the VEVs of $\xi_N$, $\chi_N$ and $\varphi_N$ are in general complex.
Apparently, $CP$ is broken in this vacuum and thus in the neutrino sector. However, as explained in subsection \ref{sec31}, such violation of $CP$ is irrelevant for (the residual symmetry of) the neutrino mass matrix $m_\nu$ and
thus the lepton mixing (at LO), as long as the VEVs $v_{\xi_N}$, $v_{\chi_N}$ and $v_{\varphi_N}$ have the same phase up to $\pi$, because the latter becomes then an overall phase which can be factored out in $m_\nu$, see 
eqs.(\ref{mnuLO},\ref{paratux}). From eq.(\ref{vNrels}) we see that this is achieved for $e_\nu/d_\nu <0$ and $(2 b_\nu e_\nu/d_\nu - c_\nu)/a_\nu >0$. Another way to show that the common phase of the VEVs $v_{\xi_N}$, $v_{\chi_N}$
and $v_{\varphi_N}$ is unphysical is to re-define the fields of the model so that this phase becomes an overall phase of the LO superpotential. 
The VEVs are expected to have the same order of magnitude that we choose to be $\lambda \, \La$, since they are directly related through the parameters of $w_{fl,\nu}$, see eq.(\ref{vNrels}).
Eventually, requesting the F-term of $\tilde{\xi}^0_N$ to vanish leads to
\begin{equation}
\label{vxiNp}
v^2_{\xi'_N} = - \frac{M^2}{a_\xi} \; .
\end{equation}
For $M^2 /a_\xi >0$ we immediately see that the VEV of $\xi'_N$ is imaginary, as it should in order to be compatible with $G_\nu$. Notice for $M \sim \lambda \, \La$ also $\langle \xi'_N \rangle$ is of this order of magnitude.
Such type of alignment has been proposed in \cite{phasesinVEVs}.

\subsection{Next-to-leading order results}
\label{sec42}

We now turn to the discussion of the non-renormalizable contributions to the flavon superpotential. These induce shifts in the vacuum obtained at LO. Taking into consideration the sizes of the LO VEVs
 (fields with an index $E$ have a VEV of order $\lambda^2 \, \La$, whereas those with an index $N$ have VEVs of order $\lambda \, \La$), we find the most relevant subleading corrections to arise from the
 following three operators
 \begin{equation}
 \label{wflNLO}
s_{\nu, 1} \, \xi_N \, (\chi_N^0 \xi'_N \chi_N)/\La +  \tilde{s}_{\nu, 1} \, \tilde{\xi}_N \, (\chi_N^0 \xi'_N \chi_N)/\La + s_{\nu, 2} \, (\varphi_N^0 \xi'_N \chi_N \varphi_N)/\La \; .
\end{equation}
Adding these to $w_{fl}$, re-computing the F-terms in a linear expansion in the shifts of the vacuum and equating them to zero, we find that  they give rise to a correction of the VEV of $\chi_N$
\begin{equation}
\delta v_{\chi_{N,1}} = -\delta v_{\chi_{N,2}} = \frac{s_{\nu,1}}{2 \, d_\nu} \, \frac{v_{\xi_N}}{\La} \, v_{\xi'_N} 
\end{equation}
with $\delta v_{\chi_{N,i}}$ denoting the shifts of the vacuum of the $i^{\rm th}$ component of $\chi_N$, $\langle \chi_{N,i} \rangle = v_{\chi_N} + \delta v_{\chi_{N,i}}$. Since $\delta v_{\chi_{N,i}}$ are
 proportional to $v_{\xi_N}$, they carry the phase that is associated with the second flat direction of the flavon potential and that is also carried by the 
 LO VEVs of the fields $\xi_N$, $\chi_N$ and $\varphi_N$. At the same time, $\delta v_{\chi_{N,i}}$ are proportional to $v_{\xi'_N}$ which is imaginary. Their relative suppression
with respect to the LO VEVs of the flavons with index $N$ is $\lambda$. Thus, the shifted vacuum of $\chi_N$ can be parameterized as
\begin{equation}
\label{VEVchiNNLO}
\langle \chi_N \rangle = v_{\chi_N} \, \left(  \begin{array}{c} 1 + i \, \alpha \, \lambda \\ 1 - i \,  \alpha \, \lambda \end{array} \right) 
\end{equation}
with $\alpha$ real. Its form leaves the group $G_\nu$ invariant (up to the above-mentioned phase of $v_{\chi_N}$), see eq.(\ref{vgnu}), because 
the operators in eq.(\ref{wflNLO}) that induce the shift only contain fields with index $N$ (especially, exactly one of them is $\xi'_N$).
The other VEVs do not acquire non-trivial shifts at this order. 
Indeed, all terms, which lead to contributions of order $\lambda^5 \, \La^2$ or larger to the F-terms of the driving fields $\xi^0_N$, $\chi^0_N$, $\varphi^0_N$ and $\tilde{\xi}^0_N$, if the LO flavon VEVs are plugged in, only comprise flavons with an 
index $N$ and thus cannot lead further non-trivial shifts in the vacuum.\footnote{Note that we have to assume here that the VEV $v_{\varphi_N}$ is real so that the LO vacuum of the flavons $\xi_N$, $\tilde{\xi}_N$, $\chi_N$ and $\varphi_N$
leaves the $CP$ symmetry intact. Otherwise shifts in the VEVs would be larger and thus also the corrections to the neutrino mass matrix.} 
The most relevant terms including the flavons $\chi_E$ and $\varphi_E$ which do change the alignment are 
\begin{equation}
\tilde{\xi}^0_N (\varphi_E^3)/\La +  \tilde{\xi}^0_N (\varphi_E^2 \chi_E)/\La + \tilde{\xi}^0_N (\chi_E^3)/\La \; .
\end{equation}
Their contribution to the F-term of $\tilde{\xi}^0_N$ is of order $\lambda^6 \, \La^2$ and hence they are relatively suppressed by $\lambda^4$ compared to the contribution from the LO terms in eq.(\ref{wflxi_LO}) (see also eq.(\ref{vxiNp}) and below). 
Since the VEVs of $\chi_E$ and $\varphi_E$ carry an undetermined phase, see eq.(\ref{vErels}), 
such contributions lead to a shift in the phase of the VEV of $\xi'_N$ so that it acquires a small real part which is suppressed by $\lambda^4$ with respect to $v_{\xi'_N}$. 

Similarly, we can discuss the subleading contributions to $w_{fl,e}$. The ones leading to shifts in the vacuum of $\chi_E$ and $\varphi_E$ that cannot be absorbed into its LO form, given in eq.(\ref{vELO}), 
comprise six flavons. Two of them belong to the
set $\{ \chi_E, \varphi_E \}$ and are necessary in order to saturate the $Z_3$ charge, while the other four ones carry an index $N$ (there are either three fields of the set $\{\xi_N, \tilde \xi_N, \varphi_N \}$ and one power of $\chi_N$ or 
three fields $\chi_N$ and one of the set $\{\xi_N, \tilde \xi_N, \varphi_N \}$). Assuming the sizes of the VEVs, as discussed above, these contributions to the F-terms of the driving fields $\xi^0_E$, $\tilde\xi^0_E$ and $\chi^0_{E}$ 
are of the order $\lambda^8 \, \La^2$ and are thus suppressed by $\lambda^4$ with respect to the contributions from the renormalizable terms in $w_{fl,e}$, see eqs.(\ref{wfle_LO},\ref{Ftermse}). Consequently, they induce shifts in the VEVs of $\chi_E$
and $\varphi_E$ at relative order $\lambda^4$  such that the latter take the form
\begin{equation}
\label{VEVsENLO}
\langle \chi_E \rangle = v_{\chi_E} \, \left( \begin{array}{c} \beta \, \lambda^4  \\ 1 \end{array} \right) \;\;\; \mbox{and} \;\;\;
\langle \varphi_E \rangle = v_{\varphi_E} \,  \left( 1 + \gamma \, \lambda^2 \right) \, \left( \begin{array}{c} \epsilon \, \lambda^4  \\ 1 \\ \eta \, \lambda^4 \end{array} \right)
\end{equation}
with $\beta$, $\gamma$, $\epsilon$ and $\eta$ having absolute values of order one and being in general complex.

\section{Summary}
\label{concl}

In this paper we have constructed a SUSY model for leptons with the flavour group $G_f=S_4 \times Z_3$ and a generalized $CP$ symmetry.
The spontaneous breaking of these symmetries to a $Z_3$ subgroup in the charged lepton and to $Z_2 \times CP (\times Z_3)$ in the neutrino sector 
leads to a mixing matrix with one free parameter. The latter is determined by the elements of the neutrino mass matrix and  
 is directly related to the reactor mixing angle. This angle is naturally of the correct order $\lambda$ in our model due to its origin from a subleading operator.
At the same time, the other mixing parameters are also predicted, especially the atmospheric mixing angle is maximal as well as the Dirac phase.
The two Majorana phases are trivial. The solar mixing angle fulfills $\sin^2 \theta_{12} \gtrsim 1/3$. 
We predict the three neutrino masses to effectively depend on only two parameters at LO and thus the mass spectrum can only be normally ordered
with the absolute neutrino mass scale being determined. 
 We have discussed in detail the vacuum  alignment of the flavour and $CP$ symmetry breaking fields and have shown that this alignment
is stable against the inclusion of higher-order terms in the flavon superpotential. 
We have computed leading as well as subleading contributions to the charged lepton and neutrino mass matrices and have shown that
all our predictions are only slightly corrected by the latter contributions. The most relevant effect of the subleading corrections is the generation
of the electron mass. Indeed, the charged lepton mass hierarchy is correctly reproduced without any tuning in our model, because the different lepton masses arise from
operators with one, two and several flavons, respectively.

It would be interesting to extend our predictive and successful model from the lepton to the quark sector, for example in the framework of a SUSY grand unified theory.

\vspace{0.3in}

{\bf Note Added:} During the completion of this manuscript another paper \cite{otherS4CP} has been submitted to the arXiv
 that contains similar SUSY models in which the flavour group $S_4$ and a generalized $CP$ symmetry are broken to $Z_2 \times CP$ in the neutrino sector and the charged lepton mass matrix is diagonal.

\section*{Acknowledgements}

FF has been partly supported by the European Programme "Unification in the LHC Era" (UNILHC, PITN-GA-2009-237920) and 
by the European Union FP7 ITN INVISIBLES (Marie Curie Actions, PITN-GA-2011-289442). CH is supported by the ERC Advanced Grant no. 267985,
"Electroweak Symmetry Breaking, Flavour and Dark Matter: One Solution for Three Mysteries" (DaMeSyFla). CH would like to thank Michael A. Schmidt for email correspondence.

\appendix
 \section{Group theory of $S_4$}
 \label{appA}

We follow the convention and notation of \cite{S4generators} and repeat for the reader's convenience the form of the $S_4$
generators $S$, $T$ and $U$ in this particular basis, the Kronecker products and the form of the Clebsch Gordan coefficients.

The generators $S$, $T$ and $U$ in the five different representations are chosen as follows
\begin{center}
\begin{math}
\begin{array}{llll}
{\bf 1}:       & S=1 \; , & T=1 \; ,  & U=1 \ ,\\[2mm]
{\bf 1^\prime}:    & S=1  \; , & T=1 \; ,  & U=-1\ ,\\[2mm]
{\bf 2}: & S= \left( \begin{array}{cc}
    1&0 \\
    0&1
    \end{array} \right) \; ,
    & T= \left( \begin{array}{cc}
    \omega&0 \\
    0&\omega^2
    \end{array} \right) \; ,
    & U=  \left( \begin{array}{cc}
    0&1 \\
    1&0
    \end{array} \right)\ ,\\[2mm]
{\bf 3}: & S= \frac{1}{3} \left(\begin{array}{ccc}
    -1& 2  & 2  \\
    2  & -1  & 2 \\
    2 & 2 & -1
    \end{array}\right) \; ,
    & T= \left( \begin{array}{ccc}
    1 & 0 & 0 \\
    0 & \omega^{2} & 0 \\
    0 & 0 & \omega
    \end{array}\right) \; ,
    & U= - \left( \begin{array}{ccc}
    1 & 0 & 0 \\
    0 & 0 & 1 \\
    0 & 1 & 0
    \end{array}\right)\ ,\\[2mm]
{\bf 3^\prime}: & S= \frac{1}{3} \left(\begin{array}{ccc}
    -1& 2  & 2  \\
    2  & -1  & 2 \\
    2 & 2 & -1
    \end{array}\right) \; ,
    & T= \left( \begin{array}{ccc}
    1 & 0 & 0 \\
    0 & \omega^{2} & 0 \\
    0 & 0 & \omega
    \end{array}\right) \; ,
    & U= \left( \begin{array}{ccc}
    1 & 0 & 0 \\
    0 & 0 & 1 \\
    0 & 1 & 0
    \end{array}\right)\ ,
\end{array}
\end{math}
\end{center}
with $\omega = e^{2\pi i/3}$.
These fulfill the relations
\begin{eqnarray}\nonumber
&& S^2 = \mathbb{1} \; , \;\; T^3 = \mathbb{1} \; , \;\; U^2= \mathbb{1} \; ,\\[0mm]
\nonumber
&&  (S T)^3 = \mathbb{1} \; , \;\; (S U)^2 = \mathbb{1} \; , \;\; (T U)^2 = \mathbb{1} \; , \;\;
  (S T U)^4 = \mathbb{1} \; .
\end{eqnarray}

In order to perform explicit computations of $S_4$ invariants, we need the Kronecker products as well as the Clebsch Gordan coefficients.
The Kronecker products are of the form
\begin{eqnarray}\nonumber
&&\bf 1 \times {\bs \mu} = {\bs \mu} \;\; \forall \;\; {\bs \mu} \; , \;\; 1^\prime \times 1^\prime =1 \; , \;\; 1^\prime \times 2 = 2 \; ,\\[0mm]
\nonumber
&&\bf 1^\prime \times 3 = 3^\prime \; , \;\; 1^\prime \times 3^\prime = 3 \; ,\\[0mm]
\nonumber
&&\bf 2 \times 2 = 1 + 1^\prime + 2 \; , \;\; 2 \times 3 = 2 \times 3^\prime = 3 + 3^\prime\; ,\\[0mm]
\nonumber
&&\bf 3 \times 3 = 3^\prime \times 3^\prime = 1 + 2 + 3 + 3^\prime \; , \;\;
3 \times 3^\prime = 1^\prime + 2 + 3 + 3^\prime \; .
\end{eqnarray}
In the following we list the Clebsch-Gordan coefficients using the notation $x
\sim {\bf   1}$, $x^\prime \sim {\bf 1^\prime}$,
$( y_1 , y_2 )^t, ( \tilde y_1, \tilde y_2 )^t \sim~{\bf 2}$,
$( z_1 ,z_2 , z_3 )^t, ( \tilde z_1 , \tilde z_2 , \tilde z_3 )^t \sim {\bf 3}$,
$( z'_1 , z'_2 , z'_3 )^t, ( \tilde z'_1 , \tilde z'_2 , \tilde z'_3 )^t \sim
{\bf 3'}$.

\noindent For a singlet multiplied with a doublet or a triplet we find
\begin{eqnarray}\nonumber
{\bf 1^{(\prime)}\times 2}: && (x y_1 , x y_2 )^t \sim {\bf 2}  \; , \;\; (x' y_1 , -x' y_2 )^t \sim {\bf 2} \; ,\\[2mm]
\nonumber
{\bf 1^{(\prime)} \times 3}: && (x z_1 , x z_2 , x z_3 )^t \sim {\bf 3} \; , \;\; (x' z_1 , x' z_2 , x' z_3 )^t \sim {\bf 3'} \; ,\\[2mm]
\nonumber
{\bf 1^{(\prime)} \times 3'}: \hspace{-1mm} && (x z'_1 , x z'_2 , x z'_3 )^t \sim {\bf 3'} \; , \;\; (x' z'_1 , x' z'_2 , x' z'_3 )^t \sim {\bf 3} \; .
\end{eqnarray}
For a doublet coupled to a doublet we have
\begin{equation}\nonumber
{\bf 2 \times 2}:\;\;\; y_1 \tilde y_2 + y_2 \tilde y_1 \sim {\bf 1} \; , \;\;
y_1 \tilde y_2 - y_2 \tilde y_1 \sim {\bf 1'} \; , \;\;
(y_2 \tilde y_2 , y_1 \tilde y_1)^t \sim {\bf 2} 
\end{equation}
and for a doublet multiplied with a triplet 
\begin{equation}\nonumber
{\bf 2 \times 3}: \;\;\,
(y_1 z_2 + y_2 z_3 , y_1 z_3 + y_2 z_1 , y_1 z_1 + y_2 z_2 )^t \sim {\bf 3} \; , \;\;
(y_1 z_2 - y_2 z_3 , y_1 z_3 - y_2 z_1 , y_1 z_1 - y_2 z_2 )^t \sim {\bf 3'} \; ,
\end{equation}
and
\begin{equation}\nonumber
{\bf 2 \times 3'}: \;\;\,
(y_1 z'_2 - y_2 z'_3 , y_1 z'_3 - y_2 z'_1 , y_1 z'_1 - y_2 z'_2 )^t \sim {\bf 3} \; , \;\;
(y_1 z'_2 + y_2 z'_3 , y_1 z'_3 + y_2 z'_1 , y_1 z'_1 + y_2 z'_2 )^t \sim {\bf
  3'}  \, .
\end{equation}
For the products ${\bf 3 \times 3}$ and ${\bf 3' \times 3'}$ (with $z_i$, $\tilde z_i$ to be replaced by $z'_i$, $\tilde z'_i$) we get
\begin{eqnarray}\nonumber
&& z_1 \tilde z_1 + z_2 \tilde z_3 +  z_3 \tilde z_2 \sim {\bf 1} \; , \;\;
(z_1 \tilde z_3 + z_2 \tilde z_2 +  z_3 \tilde z_1 , z_1 \tilde z_2 + z_2 \tilde z_1 +  z_3 \tilde z_3)^t \sim {\bf 2} \; ,\\[2mm]
\nonumber
&& (z_2 \tilde z_3 - z_3 \tilde z_2 , z_1 \tilde z_2 - z_2 \tilde z_1, z_3 \tilde z_1 - z_1 \tilde z_3)^t \sim {\bf 3} \; ,\\[2mm]
\nonumber
&& (2 \, z_1 \tilde z_1 - z_2 \tilde z_3 - z_3 \tilde z_2 , 2 \, z_3 \tilde z_3 - z_1 \tilde z_2 - z_2 \tilde z_1,
2 \, z_2 \tilde z_2 - z_1 \tilde z_3 - z_3 \tilde z_1)^t \sim {\bf 3'} \; ,
\end{eqnarray}
while the Clebsch Gordan coefficients for the product ${\bf 3 \times 3'}$ read
\begin{eqnarray}\nonumber
&& z_1 z'_1 + z_2 z'_3 +  z_3 z'_2 \sim {\bf 1'} \; , \;\;
(z_1 z'_3 + z_2 z'_2 +  z_3 z'_1 ,-( z_1 z'_2 + z_2 z'_1 +  z_3 z'_3))^t \sim {\bf 2} \; ,\\[2mm]
\nonumber
&& (2 \, z_1 z'_1 - z_2 z'_3 - z_3 z'_2 , 2 \, z_3 z'_3 - z_1 z'_2 - z_2 z'_1,
2 \, z_2 z'_2 - z_1 z'_3 - z_3 z'_1)^t \sim {\bf 3} \; ,\\[2mm]
\nonumber
&& (z_2 z'_3 - z_3 z'_2 , z_1 z'_2 - z_2 z'_1, z_3 z'_1 - z_1 z'_3)^t \sim {\bf 3'} \; .
\end{eqnarray}


\end{document}